\documentclass[12pt,thmsa]{article}
\usepackage{sw20rui}

\input{tcilatex}
\begin{document}

\author{Ezra T. Newman \\
Dept of Physics and Astronomy, University \\
of Pittsburgh, Pittsburgh PA 15260}
\title{On a Classical, Geometric Origin of Magnetic Moments, Spin-Angular Momentum
and the Dirac Gyromagnetic Ratio }
\date{Jan. 16, 2001 }
\maketitle

\begin{abstract}
By treating the real Maxwell Field and real linearized Einstein equations as
being imbedded in complex Minkowski space, one can \textit{interpret }%
magnetic moments and spin-angular momentum as arising from a charge and mass 
\textit{monopole} source moving along a complex world line in the complex
Minkowski space. In the circumstances where the \textit{complex} center of
mass world-line coincides with the \textit{complex} center of charge
world-line, the gyromagnetic ratio is that of the Dirac electron.
\end{abstract}

\section{Introduction}

Many years ago in a brief note\cite{NW} concerning relativistic angular
momentum, we made the claim, largely as an afterthought and with only a
skimpy argument presented, that there was a simple\thinspace geometric
classical argument that led to the Dirac value of the gyromagnetic ratio of
a charged massive particle. There was no mention of either quantum mechanics
or the Dirac equation. The argument made then, was based on a kinematic
analogy concerning certain algebraic properties of the angular momentum
tensor and the electric-magnetic dipole tensor, and has on several occasions
been reasonably questioned. In the present note we return to the same issue
and give the complete argument, including details of the dynamics that lead
to the relevant algebraic properties.

\qquad We make the following pair of claims: Consider first both the vacuum
Maxwell field and the linearized vacuum gravitational field on (ordinary
real) Minkowski space. It is possible to view (purely in a mathematical
sense by the analytic continuation of the real fields into the complex) that
the magnetic moment $\overrightarrow{\mu }$ of a particle with charge $q$,
arises from the particle being displaced an imaginary distance ($i%
\overrightarrow{Y}_{em}$) into complex Minkowski space so that $%
\overrightarrow{\mu }=q\overrightarrow{Y}_{em}.$ Also, classical
relativistic spin-angular momentum $\overrightarrow{S},$ can be interpreted
(using linearized general relativity) as arising from the displacement ($i%
\overrightarrow{Y}_{g}$) of a massive monopole particle, $m,$ (Schwarzschild
solution) also into the complex so that $\overrightarrow{S}=mc%
\overrightarrow{Y}_{g}.$ We refer to $\overrightarrow{Y}_{em}$ and $%
\overrightarrow{Y}_{g}$ as respectively, the complex centers of charge and
mass. The second claim is simply the observation that when the two imaginary
displacements are the same, i.e., $\overrightarrow{Y}_{em}=\overrightarrow{Y}
_{g},$ then the gyromagnetic ratio \TEXTsymbol{\vert}$\mu |/|S|$ is the
Dirac value, $q/mc.$ This can be restated more geometrically as; given a
real solution of the linearized Einstein equations (on real Minkowski space)
with non-vanishing mass and a solution of the Maxwell equations with
non-vanishing charge, if the associated angular momentum and magnetic moment
are \textit{interpreted} as arising from a \textit{complex} \textit{center
of mass} and a \textit{complex center of charge} \textit{that coincide},
then the resulting gyromagnetic ration is that of Dirac. We stress that this
is simply an observation and it remains to be shown if it has any further
physical significance.

A point that we want to emphasis is that we are dealing, here, strictly with
the vacuum fields and are not considering the nature or structure of the
sources. What we are calling the charge, the dipole moments, the mass and
angular momentum are the coefficients of the relevant terms in $r^{-1}$
expansions of the Maxwell and gravitational fields. The structure of the
sources has been discussed elsewhere\cite{JN,WI1,WI2,B,GK} .

Another point that we wish to make is that this result, though it appears to
depend on the linearized Einstein equations, remains though in a different
form in full GR. The charged Kerr solution (with parameters $(m,q,Y)$ also
has the Dirac gyromagnetic ratio and can be ``obtained'' from the
Schwarzschild solution by a complex translation\cite{JN}.

In Sec.II, we will write the Maxwell equations in such a way so that the
complex interpretation becomes obvious and furthermore show how the real
simple Coulomb solution, by a complex translation, can be reinterpreted as
having a magnetic moment. In Sec. III, we will rewrite the linearized vacuum
Einstein equations as equations for the Weyl tensor (often considered as the
gravitational analogue of the Maxwell tensor) and see its remarkable
similarity with the Maxwell equations. The complex translation
interpretation of angular momentum simply follows as does our conclusion
concerning the gyromagnetic ratio. Finally, in Sec. IV, we point out
possible physical consequences of these observations. In an appendix we
examine, without a real resolution, how a rather strange complex coordinate
transformation is equivalent to a real transformation.

\section{The Vacuum Maxwell Equations}

We begin with the standard vacuum Maxwell equations defined on Minkowski
space,

\begin{eqnarray}
\nabla _{a}F^{ab} &=&0,  \label{max1} \\
\nabla _{[a}F_{cd]} &=&0\Leftrightarrow \nabla _{a}F^{*ab}=0  \label{max2}
\end{eqnarray}
with 
\begin{eqnarray}
F^{*ab} &=&\frac{1}{2}\eta ^{abcd}F_{cd},  \label{dual1} \\
F^{**ab} &=&-F^{ab}.  \label{dual2}
\end{eqnarray}

We use for the flat metric $\eta ^{ab}=diag(1,-1,-1,-1)$ and $\eta
_{0123}=-\eta ^{0123}=1,$ with $\eta _{0ijk}=\epsilon _{ijk}$ and $\eta
^{0ijk}=-\epsilon ^{ijk}.$

By defining the (self-dual) complex combination

\begin{eqnarray}
W^{ab} &\equiv &F^{ab}-iF^{*ab},  \label{W} \\
W^{*ab} &=&F^{*ab}+iF^{ab}=iW^{ab}  \label{SD1} \\
W^{ab} &=&-iW^{ab*},  \label{W*}
\end{eqnarray}

the Maxwell equations become the single complex equation 
\begin{equation}
\nabla _{a}W^{ab}=0.  \label{complexmax}
\end{equation}

\begin{remark}
It is often useful to see the matrix form of $W^{ab}$ in terms of Cartesian
components of the electric and magnetic fields, i.e., as

$W^{ab}=F^{ab}-iF^{*ab}=$

$\left( 
\begin{array}{llll}
0, & -(E^{1}+iB^{1}), & -(E^{2}+iB^{2}), & -(E^{3}+iB^{3}) \\ 
E^{1}+iB^{1}, & 0, & -i(E^{3}+iB^{3}), & i(E^{2}+iB^{2}) \\ 
E^{2}+iB^{2}, & i(E^{3}+iB^{3}), & 0, & -i(E^{1}+iB^{1}) \\ 
E^{3}+iB^{3}, & -i(E^{2}+iB^{2}), & i(E^{1}+iB^{1}), & 0
\end{array}
\right) $
\end{remark}

By decomposing Eq.(\ref{complexmax}), into its time-space parts, using $%
(i,j,k)$ for the space part and $0$ for the time part and

\begin{eqnarray}
W^{ij} &=&i\epsilon ^{ijk}W_{0k}=-i\epsilon ^{ijk}W^{0k},
\label{decomposition} \\
W^{i} &=&W^{0i}  \nonumber
\end{eqnarray}
we have

\begin{eqnarray}
\nabla _{0}W^{0j}+i\epsilon ^{jik}\nabla _{i}W^{0k} &=&0
\label{space-timemax} \\
\nabla _{i}W^{0i} &=&0.  \nonumber
\end{eqnarray}

In 3-vector terms, it becomes

\begin{eqnarray}
\partial _{t}\overrightarrow{W}+i\,curl\,\overrightarrow{W} &=&0  \label{3+1}
\\
div\,\overrightarrow{W} &=&0  \nonumber
\end{eqnarray}
the vector form of Maxwell's equations, with 
\begin{equation}
\overrightarrow{W}=\overrightarrow{E}+i\overrightarrow{B}.  \label{Wv}
\end{equation}

The value in this formulation, is that we can now think of the equations, as
equations for a complex vector on a four complex dimensional manifold, $%
M_{C},$ the complexification of Minkowski space, $M$.

When any complex solution is found as a function of the four complex
coordinates, 
\begin{equation}
z^{a}=x^{a}+iy^{a}  \label{clxcoor}
\end{equation}
i.e., of the form, 
\begin{equation}
\overrightarrow{W}(x^{a}+iy^{a})
\end{equation}
one can simply chose, $y^{a}=c^{a},$ as four real constants and then take
the real and imaginary parts of $\overrightarrow{W}$ to construct an$%
\overrightarrow{E}$ and $\overrightarrow{B},$ i.e., a real Maxwell field on
real $M.$ Since the equations do not have any explicit coordinate
dependence, solutions will go over to solutions if they are transformed by
complex Poincare transformations. If the transformations were real then the
resulting new solutions are completely equivalent to the untransformed
solutions; but if the transformations are intrinsically complex (or
imaginary) then the new real solutions are genuinely different from the
starting solution.

We will illustrate this with the Coulomb solution (in the vacuum region)
transformed by a complex translation\cite{N} and in the process show how the
magnetic moment arises from the complex translation.

Considering the Coulomb solution in any of the following forms

\begin{eqnarray}
E^{i} &=&qr^{-2}n^{i},\text{ }B^{i}=0\Longleftrightarrow W^{i}=er^{-2}n^{i}
\label{coulomb} \\
F^{ab} &=&\frac{2qn^{[a}v^{b]}}{r^{2}}  \nonumber
\end{eqnarray}
with a unit radial vector, $n^{i}=\frac{x^{i}}{r}$ and particle velocity $%
v^{\nu }=(1,0,0,0)$ given in the rest-frame, we study the ``same solution''
thought of as translated into the complex, i.e., we replace the real
coordinates by the complex translated ones;

\begin{equation}
x^{i}\rightarrow \widehat{x}^{i}=x^{i}-C^{i}  \label{CompTrans}
\end{equation}
with

\begin{equation}
C^{i}=X^{i}+iY^{i}  \label{Ci}
\end{equation}
a complex spatial translation. If we denote the complex translated
quantities with a hat (\symbol{94}) then, keeping only low powers of $%
\widehat{r}^{-1},$ we have

$\qquad \widehat{r}^{2}=(x^{i}-C^{i})^{2}=x^{i2}-2x^{i}C^{i}+C^{i}\cdot
C^{i}=r^{2}(1-2n^{i}\frac{C^{i}}{r}+O(r^{-2}))\qquad $

$\qquad \widehat{r}=r(1-r^{-1}n\cdot C+..)$

$\qquad \widehat{r}^{-2}=r^{-2}(1+2r^{-1}n\cdot C+...)$

\begin{mathletters}
\begin{equation}
\widehat{n}^{i}=\frac{\widehat{x}^{i}}{\widehat{r}}=\frac{x^{i}-C^{i}}{
r(1-r^{-1}n\cdot C)}=\frac{(x^{i}-C^{i})}{r}(1+r^{-1}n\cdot C....)
\label{nhat}
\end{equation}

The new solution, obtained from the complex translated space coordinates,
keeping terms up to $r^{-3},$ is

$\qquad $%
\end{mathletters}
\begin{eqnarray}
\widehat{W}^{i} &=&q\widehat{r}^{-2}\widehat{n}^{i}, \\
\widehat{W}^{i} &=&qr^{-2}(n^{i}-\frac{C^{i}}{r})(1+2r^{-1}n\cdot
C)(1+r^{-1}n\cdot C), \\
\widehat{W}^{i} &=&qr^{-2}\{n^{i}-\frac{C^{i}}{r}+3n^{i}r^{-1}n\cdot C\}
\end{eqnarray}

$\qquad $

\begin{equation}
\widehat{W}^{i}=qr^{-2}n^{i}+r^{-3}\{3n^{i}n\cdot
D_{(C)}-D_{(C)}^{i}\}+O(r^{-4})  \label{What}
\end{equation}
with the complex dipole moment $D_{(C)}^{i}$

\begin{equation}
D_{(C)}^{i}=qC^{i}\text{ and }n\cdot D_{(C)}=D_{(C)}^{i}n^{i}.  \label{D}
\end{equation}

This corresponds\cite{L&L} to a static electric monopole and dipole field
with, in addition, a magnetic dipole field. The real part of $D^{i}$ is the
electric dipole moment,

\begin{equation}
\mathit{Re}D^{i}=qX^{i},
\end{equation}
and the imaginary part is the magnetic dipole moment,

\begin{equation}
\mathit{\func{Im}}D^{i}=\mu ^{i}=qY^{i};  \label{mu}
\end{equation}
\textit{both} arising from the complex translation. Note that identifying
the magnetic moment with $qY^{i}$ is dimensionally correct.

The relativistic version of Eq.(\ref{What}) is

\begin{equation}
\widehat{W}^{ab}=2qr^{-2}n^{[a}v^{b]}-r^{-3}%
\{2n^{[a}v^{b]}3D_{(C)}^{ab}n_{a}v_{b}+D_{(C)}^{ab}\}  \label{relWhat}
\end{equation}
with the complex dipole moment

\begin{eqnarray}
D_{(C)}^{ab} &=&2q(X^{[a}+iY^{[a})v^{b]},  \label{dipoletensor} \\
X^{b}v_{b} &=&Y^{b}v_{b}=0,
\end{eqnarray}
and

\begin{equation}
n\cdot D_{(C)}=-D_{(C)}^{ab}n_{a}v_{b},
\end{equation}
remembering that $n_{a}n_{b}\eta ^{ab}=-1.$

If we began with simply a moving charge, $q,$ with an electric dipole moment 
$D_{(e)}^{a}$ and magnetic dipole moment $D_{(m)}^{a},$ such that $%
D_{(m)}^{a}v_{a}=D_{(e)}^{a}v_{a}=0,$ $v^{a}$ being the particle
four-velocity a real dipole tensor can be defined by $%
D_{(R)}^{ab}=2D_{(e)}^{[a}v^{b]}+\eta ^{abcd}D_{(m)c}v_{d}.$ Under a shift
in origin, $\widehat{x}^{a}=x^{a}-X^{a},$ it transforms as

\begin{equation}
D_{(R)}^{\prime \,ab}=D_{(R)}^{ab}-2qX^{[a}v^{b]}  \label{D'}
\end{equation}
so that the electric part of $D_{(R)}^{ab}$ can be transformed away, with
arbitrary $\lambda ,$ by 
\begin{equation}
X^{a}=q^{-1}D_{(e)}^{a}+v^{a}\lambda .  \label{CofC}
\end{equation}
This defines the (real) center of charge world line.

In an analogous fashion, from the complex point of view, the entire complex
dipole moment can be transformed away, (in the stationary case) by

\begin{equation}
D_{(C)}^{\prime \,ab}=D_{(C)}^{ab}-2qZ^{[a}v^{b]}=0  \label{CD}
\end{equation}
with 
\begin{equation}
Z^{a}=C^{a}+v^{a}\lambda \equiv (X^{a}+iY^{a})+v^{a}\lambda ,  \label{CCofC}
\end{equation}
defining (with arbitrary complex $\lambda )$ the \textit{complex center of
charge }world line. In the time dependent case neither dipole can be
transformed way but one could still define the complex center of charge so
that instantaneously in some frame it can be transformed away.

Viewed in this formal manner, we see that both an electric and magnetic
dipole can be interpreted as arising from a charged particle (with no dipole
moments) moving along a complex world line.

\section{Linearized General Relativity}

Though in general relativity, (GR), the most common point of view is to
think of the metric tensor as the basic and physical field variable with
other variables obtained from the metric, there are nevertheless good
reasons to consider other variables as perhaps more fundamental. In
particular, one often thinks of the metric as analogous the vector potential
of Maxwell theory with the Weyl tensor the analogue of the Maxwell field.
Without being ideological about it, we will see that, at least, in linear
theory the analogy is remarkably accurate.

Before we consider the linearization of GR, several remarks about the Weyl
tensor are useful. The Weyl tensor, $C_{abcd},$ which is the trace-free part
of the full curvature tensor is equal to the curvature tensor when the
vacuum Einstein equations are satisfied - with no cosmological constant.
Under these circumstance the Bianchi Identities 
\begin{equation}
R_{ab[cd:e]}\equiv 0  \label{BI}
\end{equation}
become differential equation for the Weyl tensor,

\begin{equation}
\nabla _{[e}C_{cd]ab}=0  \label{WBI}
\end{equation}
or 
\begin{equation}
\nabla _{c}C^{*cdab}=0  \label{DWBI}
\end{equation}
where the dual on the first pair is

\begin{equation}
C^{*cdab}=\frac{1}{2}\eta ^{cdef}C_{ef}^{\quad ab}.  \label{DW}
\end{equation}
Since dualing on the right pair yields\cite{PR} 
\begin{equation}
C^{*cd*ab}=-C^{cdab}  \label{DD}
\end{equation}
Eq.(\ref{DWBI}) can be written as

\begin{equation}
\nabla _{c}C^{cdab}=0.
\end{equation}
By combining them and relabeling indices, we obtain, using the complex Weyl
tensor, our basic field, the self-dual Weyl tensor 
\begin{equation}
W^{abcd}=C^{abcd}-iC^{*abcd}  \label{CW}
\end{equation}
with its field equation 
\begin{equation}
\nabla _{a}W^{abcd}=0.  \label{DBI}
\end{equation}
Note that

\begin{equation}
W^{abcd*}=iW^{abcd}.  \label{SD2}
\end{equation}

By linearized GR, we will mean Eq.(\ref{DBI}), where the covariant
derivative is taken with the flat metric connection. This is the first
example of the analogy with Maxwell theory, via Eq.(\ref{complexmax}).

In analogy with the fact that all the components of $W^{ab}$ can be
expressed in terms of $W^{0i},$ we have the following identities coming from
Eq.(\ref{SD2})

\begin{eqnarray}
W_{ef}^{\quad ij} &=&-i\eta ^{ijk0}W_{efk0}=-i\epsilon ^{ijk}W_{efk0}
\label{Identities} \\
W_{lm}^{\quad ij} &=&-i\epsilon ^{ijk}W_{lmk0}=i\epsilon ^{ijk}\epsilon
_{lmn}W_{0k}^{\quad n0}  \nonumber \\
W^{ij0k} &=&-i\epsilon ^{ijl}W_{\quad l0}^{0k}=i\epsilon ^{ijl}W^{0kl0} 
\nonumber
\end{eqnarray}
which allows all components of $W^{abcd}$ to be expressed in terms of

\begin{equation}
Z^{ij}=W^{0i0j}=C^{0i0j}-iC^{*0i0j}.  \label{Z}
\end{equation}
$Z^{ij}$ is a complex, symmetric, trace-free, 3x3, tensor that contains all
the 10 components of the Weyl tensor.

Writing this out in 3+1 notation, we first have

\begin{equation}
\nabla _{0}W^{0bcd}+\nabla _{i}W^{ibcd}=0
\end{equation}
which, with $b=j$ and $b=0,$ further decomposes into

\begin{eqnarray}
\nabla _{0}W^{0jcd}+\nabla _{i}W^{ijcd} &=&0  \label{b=j} \\
\text{\qquad }\nabla _{i}W^{i0cd} &=&0.  \label{b=0}
\end{eqnarray}
With $cd=0k$ and $cd=kl$ we have

\begin{eqnarray}
\text{ }\nabla _{0}W^{0j0k}+\nabla _{i}W^{ij0k} &=&0  \label{cd=0k} \\
\nabla _{0}W^{0jkl}+\nabla _{i}W^{ijkl} &=&0  \label{cd=kl}
\end{eqnarray}
and 
\begin{eqnarray}
\nabla _{i}W^{i0j0} &=&0  \label{j0} \\
\nabla _{i}W^{i0jk} &=&0.  \label{jk}
\end{eqnarray}
Using the Identities, (\ref{Identities}), the last equation from both
previous sets follow from the first equation of the set.

We thus have that $\nabla _{a}W^{abcd}=0$ is equivalent to

\begin{eqnarray}
\nabla _{0}W^{0i0k}+i\epsilon ^{ijl}\nabla _{j}W^{0l0k} &=&0
\label{Einstein} \\
\nabla _{i}W^{i0j0} &=&0  \nonumber
\end{eqnarray}
or, with $Z^{ij}=E^{ij}+iB^{ij}=W^{0i0j}$ in dyadic form, $%
\overleftrightarrow{Z}=\overleftrightarrow{E}+i\overleftrightarrow{B},$

\begin{eqnarray}
\partial _{t}\overleftrightarrow{Z}+i\cdot curl\overleftrightarrow{Z} &=&0
\label{3+1gr} \\
div\overleftrightarrow{Z} &=&0.  \nonumber
\end{eqnarray}

These are the linearized Einstein Equations for the Weyl tensor. Note the
extraordinary similarity in form, with the complex version of the Maxwell
Equations, (\ref{3+1}). Again, as in the Maxwell case, we can generate new
solutions, by complex Poincare transformations, from old solutions or
equivalently by finding complex solutions depending on the complex Minkowski
coordinates, $z^{a}=x^{a}+iy^{a}$ and choosing $y^{a}$ as constants.

\begin{remark}
It has been well known for many years that Eq.(\ref{DBI}) can be viewed as
the linear Einstein equations. We, however, have not been able to find in
the literature the particular form, (\ref{3+1gr}), that so mimics the
Maxwell equations. Though we would be surprised, it might well be new.
\end{remark}

We apply this idea to the monopole (or linearized Schwarzschild solution)
and obtain the linearized Kerr solution. Beginning with the monopole
solution in any of the equivalent forms

\begin{eqnarray}
W_{\ }^{acbd} &=&3mr^{-3}(v^{[a}n^{c]}v^{[b}n^{d]}+\frac{1}{3}4v^{[a}\eta
^{c][d}v^{b]}),  \label{monopole} \\
Z_{\ }^{ij} &=&W_{\ }^{0i0j}=3mr^{-3}(n^{i}n^{j}-\frac{1}{3}\delta ^{ij}), 
\nonumber \\
\Psi _{2} &=&\frac{1}{2}Z^{ij}n^{i}n^{j}=m/r^{3},  \nonumber
\end{eqnarray}
with mass $m,$ velocity vector $v^{a}=\delta _{0}^{a}$ and unit radial
vector $n^{i}=\frac{x^{i}}{r},(n^{a}n_{a}=-1)$ we apply the complex
translation 
\begin{equation}
x^{i}\rightarrow \widehat{x}^{i}=x^{i}-C^{i}=C^{i}=x^{i}-(X^{i}+iY^{i}).
\label{c.t.}
\end{equation}
The new solution is first written as

\begin{mathletters}
\begin{equation}
\widehat{Z}^{ij}=3m\widehat{r}^{-3}(\widehat{n}^{i}\widehat{n}^{j}-\frac{1}{3%
}\delta ^{ij})  \label{Zhat}
\end{equation}
with (\textit{ignoring terms of higher order in} $r^{-1})$%
\end{mathletters}
\begin{eqnarray}
\widehat{r}^{2} &=&\widehat{x}^{i}\cdot \widehat{x}%
^{i}=(x^{i}-C^{i})^{2}=x^{i}\cdot x^{i}-2x^{i}\cdot C^{i}+C^{i}\cdot C^{i}
\label{r2hat} \\
&=&r^{2}-2x^{i}C^{i}+C^{i2}=r^{2}(1-2r^{-1}n\cdot C)  \nonumber \\
\widehat{r} &=&r(1-r^{-1}n\cdot C)  \nonumber \\
\widehat{r}^{-3} &=&r^{-3}(1+3r^{-1}n\cdot C)  \nonumber
\end{eqnarray}
and

\begin{eqnarray}
\widehat{n}^{i} &=&\frac{\widehat{x}^{i}}{\widehat{r}}  \label{nhat2} \\
\widehat{n}^{i} &=&\widehat{r}^{-1}(x^{i}-C^{i})=(1+r^{-1}\overrightarrow{n}
\cdot \overrightarrow{C}+..)(n^{i}-r^{-1}C^{i})  \nonumber \\
\widehat{n}^{i} &=&n^{i}+n^{i}r^{-1}\overrightarrow{n}\cdot \overrightarrow{C%
}-r^{-1}C^{i}  \nonumber
\end{eqnarray}

Substituting (\ref{nhat2}) and (\ref{r2hat}) into (\ref{Zhat}), by
expanding, collecting and keeping only terms linear in $r^{-1}C^{i},$ we have

\begin{eqnarray}
\widehat{Z}^{ij} &=&3m\widehat{r}^{-3}(\widehat{n}^{i}\widehat{n}^{j}-\frac{1%
}{3}\delta ^{ij})  \label{Zkerr} \\
&=&3mr^{-3}(1+3r^{-1}\overrightarrow{n}\cdot \overrightarrow{C})\cdot 
\nonumber \\
&&\{(n^{i}+n^{i}r^{-1}\overrightarrow{n}\cdot \overrightarrow{C}
-r^{-1}C^{i})(n^{j}+n^{j}r^{-1}\overrightarrow{n}\cdot \overrightarrow{C}
-r^{-1}C^{j}) \\
&&-\frac{1}{3}\delta ^{ij}\} \\
&=&3mr^{-3}(n^{i}n^{j}-\frac{1}{3}\delta ^{ij})  \nonumber \\
&&+3mr^{-4}\{\overrightarrow{n}\cdot \overrightarrow{C}(5n^{i}n^{j}-\delta
^{ij})-(n^{i}C^{j}+n^{j}C^{i})\}
\end{eqnarray}
or 
\begin{eqnarray}
\widehat{Z}^{ij} &=&3mr^{-3}(n^{i}n^{j}-\frac{1}{3}\delta ^{ij})
\label{Zkerr2} \\
&&+3r^{-4}\{\overrightarrow{n}\cdot \overrightarrow{D}_{(g)}(5n^{i}n^{j}-%
\delta ^{ij})-(n^{i}D_{(g)}^{j}+n^{j}D_{(g)}^{i})\}  \nonumber
\end{eqnarray}
with $D_{(g)}^{k}=mC^{k}=m(X^{k}+iY^{k}),$ the complex gravitational dipole
moment. This result is the analogue of the Maxwell result, (\ref{What}). $%
ReD_{(g)}^{k}=mX^{k}$ is the mass dipole moment and, as we will see shortly, 
$ImD_{(g)}^{k}=mY^{k}$ is the spin-angular momentum.

\begin{remark}
The quantity $\Psi _{2}=\frac{1}{2}\widehat{Z}^{ij}n^{i}n^{j}=mr^{-3}+3r^{-4}%
\overrightarrow{n}\cdot \overrightarrow{D}_{(g)}$ obtained from (\ref{Zkerr2}%
) is the same as the linearized Kerr solution. In that case the Kerr
parameter $a,$ the angular momentum per unit mass, is the same as the
magnitude of our $Y^{i}.$
\end{remark}

The relativistic version of $D_{(g)}^{k}$ is given by

\begin{equation}
D_{(g)}^{ab}=2D_{(g)}^{[a}v^{b]}  \label{Dab}
\end{equation}

with $v_{a}$ the particle velocity and

\begin{eqnarray}
D_{(g)}^{a} &=&D_{(g)}^{ab}v_{b},  \label{Dg} \\
\overrightarrow{n}\cdot \overrightarrow{D}_{(g)} &=&D_{(g)}^{ab}n_{b}v_{a}, 
\nonumber \\
D_{(g)}^{a}v_{a} &=&0.  \nonumber
\end{eqnarray}

\qquad When $v_{a}=(1,0,0,0),$ then $D_{(g)}^{a}=(0,D_{(g)}^{k}).$

Note that we can identify (and decompose) $D_{(g)}^{ab}$ with respect to a
real tensor $M^{ab}$ (the angular momentum tensor) by

\begin{equation}
D_{(g)}^{ab}=(M^{ab}-iM^{*ab})  \label{identify}
\end{equation}
where $M^{ab}$ can be further decomposed into 
\begin{equation}
M^{ab}=L^{ab}+S^{ab}.  \label{angmom}
\end{equation}
so that $L^{ab}$ (the orbital angular momentum tensor) is defined by

\begin{equation}
L^{ab}=m2X^{[a}v^{b]}=2X^{[a}p^{b]}  \label{orbital}
\end{equation}
with $X^{a}$ obtained from the $M^{ab}$ by 
\begin{equation}
mX^{a}=M^{ab}v_{b}=m^{-1}M^{ab}p_{b}
\end{equation}
and

\begin{equation}
S^{*ab}=-2S^{[a}v^{b]}.
\end{equation}
where $S^{b},$ the scaled Pauli-Lubanski spin vector, is obtained from $%
M^{ab}$ by 
\begin{equation}
S^{b}=v_{a}M^{*ab}=m^{-1}p_{a}M^{*ab}.  \label{Pauli}
\end{equation}
In the rest frame of the source, $v_{a}=(1,0,0,0),$ we have that

$M^{ab}=\left( 
\begin{array}{llll}
0, & -mX_{x}, & -mX_{y}, & -mX_{z} \\ 
mX_{x}, & 0, & -S_{z}, & S_{y} \\ 
mX_{y}, & S_{z}, & 0, & -S_{x} \\ 
mX_{z}, & -S_{y}, & S_{x}, & 0
\end{array}
\right) $

$\qquad =\left( 
\begin{array}{llll}
0, & -D_{x}, & -D_{y}, & -D_{z} \\ 
D_{x}, & 0, & -S_{z}, & S_{y} \\ 
D_{y}, & S_{z}, & 0, & -S_{x} \\ 
D_{z}, & -S_{y}, & S_{x}, & 0
\end{array}
\right) $

with $D^{i}=mX^{i},$ so that

\begin{eqnarray}
&&(M^{ab}-iM^{*ab})  \nonumber \\
&=&\left( 
\begin{array}{llll}
0, & -(mX_{x}+iS_{x}), & -(mX_{y}+iS_{y}), & -(mX_{z}+iS_{z}) \\ 
mX_{x}+iS_{x}, & 0, & i(mX_{z}+iS_{z}), & -i(mX_{y}+iS_{y}) \\ 
mX_{y}+iS_{y}, & -i(mX_{z}+iS_{z}), & 0, & i(mX_{x}+iS_{x}) \\ 
mX_{z}+iS_{z}, & i(mX_{y}+iS_{y}), & -i(mX_{x}+iS_{x}), & 0
\end{array}
\right) .
\end{eqnarray}

Restoring the $c,$ that has been tacitly taken as $1,$ we have the
dimensionally correct expression

$(M^{ab}-iM^{*ab})=$

\noindent $\left( 
\begin{array}{llll}
0, & -(mX_{x}+ic^{-1}S_{x}), & -(mX_{y}+ic^{-1}S_{y}), & 
-(mX_{z}+ic^{-1}S_{z}) \\ 
mX_{x}+ic^{-1}S_{x}, & 0, & i(mX_{z}+ic^{-1}S_{z}), & -i(mX_{y}+ic^{-1}S_{y})
\\ 
mX_{y}+ic^{-1}S_{y}, & -i(mX_{z}+ic^{-1}S_{z}), & 0, & i(mX_{x}+ic^{-1}S_{x})
\\ 
mX_{z}+ic^{-1}S_{z}, & i(mX_{y}+ic^{-1}S_{y}), & -i(mX_{x}+ic^{-1}S_{x}), & 0
\end{array}
\right) .$

\noindent which, with (\ref{Dab}) and (\ref{identify}), allows us to
identify the spin angular momentum of the source with the complex
displacement, i.e., 
\begin{equation}
S^{k}=mcY^{k}.  \label{spin}
\end{equation}

If, in Minkowski space, we have a massive charged particle at rest at the
spatial origin, producing both an electric monopole field and mass monopole
field and then (formally) consider this particle moved into the complex by
an amount $C^{i}=iY^{i},$ the new real solution has now both a magnetic
dipole moment, (\ref{mu}), $\mu ^{i}=qY^{i}$ and a spin angular moment, $%
S^{i}=cmY^{i}$. From this it follows that 
\begin{equation}
\mu ^{i}=\frac{q}{mc}S^{i}  \label{gyro1}
\end{equation}
and hence 
\begin{equation}
g_{e}=\frac{q}{mc}  \label{gyro2}
\end{equation}
obtaining the Dirac value of the gyromagnetic ratio by a purely classical
argument.

\section{Conclusions}

It has long been known, first pointed out by Brandon Carter, that the exact
charged Kerr\cite{ChKerr} solution of the Einstein-Maxwell equations
possessed the Dirac gyromagnetic ratio. Though at first this created a bit
of a stir and interest, it nevertheless did not go anywhere and eventually
faded from general interest. Nevertheless, in one form or another,
intermittently there have been a long series of papers related to the issue.
In some sense the present paper appears to give a simple geometric origin
for Carter's observation. First it suggests that a particle's magnetic
moment and intrinsic spin arises as a ``shadow'' or ``projection'' into the
real, of a particle moving in the complex space-time and that in particular
the Dirac value arises when a particle's complex center of charge \textit{\
coincides} with its complex center of mass. One's first hope is that there
is something profound about this result and perhaps there is, nevertheless
after much thought, we and others have not seen anyway to further develop
these ideas and at the present the issues lie dormant\cite{Pfister}.
Nevertheless, we are led to conjecture that every massive, charged
elementary particle with spin will possess the Dirac gyromagnetic ratio.

As a final item, we point out that though the example used here was based on
the static electric and mass monopole fields, these can be greatly
generalized. One can construct real Maxwell and linearized Einstein fields
by considering the generalization of the Lienard-Wiechart fields to that of
a point source moving along an arbitrary complex world-line in complex
Minkowski space. This leads to real fields with time varying magnetic
dipoles and spin-angular momentum. The extension of these ideas to the full
non-linear Einstein-Maxwell equations leads to algebraically special metrics%
\cite{newman-lind}. Considerable work is still required to fully understand
them.

\section{Acknowledgments}

We sincerely thank both Gerry Kaiser and Andrzej Trautman for their
questions, proddings, criticisms and insights that led to this work. This
work was supported by the NSF under grant \#PHY-0088951.

\section{Appendix}

Close 35 years ago we found a strange ``derivation\cite{JN}'' of the Kerr
metric beginning with the Schwarzschild metric, by a rather mysterious
process (trick!) involving complex transformations. \{The \textit{charged}
Kerr metric was first found\cite{ChKerr}, from the Reisner-Nordstrom metric,
in this manner.\} Up to the present there does not yet appear to be a full
rational explanation for the trick, except that it worked. Here we will give
a partial explanation but confined to the flat-space limit. The explanation
we offer is given via the complex translations, Eq.(\ref{CompTrans}), that
were discussed earlier.

First reviewing the process, we began with the Schwarzschild metric
expressed in terms of a null tetrad, i.e., 
\begin{equation}
g_{(S)}^{ab}=l^{a}n^{b}+n^{a}l^{b}-m^{a}\overline{m}^{b}-\overline{m}%
^{a}m^{b}  \label{R-N}
\end{equation}
with 
\begin{eqnarray}
l^{a}\partial _{a} &=&\partial _{r}  \label{tetrad} \\
n^{a}\partial _{a} &=&\partial _{u}-(\frac{1}{2}-\frac{M}{r})\partial _{r} 
\nonumber \\
m^{a}\partial _{a} &=&\frac{1}{\sqrt{2}r}(\partial _{\theta }+\frac{i}{\sin
\theta }\partial _{\varphi }).  \nonumber
\end{eqnarray}

Then by considering $r$ to be complex, (i.e., $r\Rightarrow r_{c})$ and
performing the complex coordinate transformation

\begin{eqnarray}
r_{c} &=&r^{*}+ia\cos \theta ^{*}  \label{complex-transformation1} \\
u_{c} &=&u^{*}-ia\cos \theta ^{*}  \label{complex transformation2} \\
\theta &=&\theta ^{*},\quad \varphi =\varphi ^{*}  \label{complex3}
\end{eqnarray}
on the tetrad vectors, yielding $l^{*a}\partial _{a},$ $n^{*a}\partial _{a},$
$m^{*a}\partial _{a},$ and $\overline{m}^{*a}\partial _{a}$ (with $\overline{%
m}^{*}$ the complex conjugate of $m^{*})$ the new metric, 
\begin{equation}
g_{(K)}^{*ab}=l^{*a}n^{*b}+n^{*a}l^{*b}-m^{*a}\overline{m}^{*b}-\overline{m}
^{*a}m^{^{*}b}  \label{ChK}
\end{equation}
depending on two parameters, $(M,a),$ automatically satisfied the Einstein
equations and was the known the Kerr metric. When the mass $(M)$ vanishes
the metric is flat but is expressed in the unusual ``Kerr coordinate
system'' and has the form;

\QTP{Body Math}
\begin{eqnarray}
ds^{2} &=&du^{*2}+2du^{*}dr^{*}-2a\sin ^{2}\theta ^{*}dr^{*}d\varphi ^{*}
\label{flat1} \\
&&-(r^{*2}+a^{2}\cos ^{2}\theta ^{*})d\theta ^{*2}-(a^{2}+r^{*2})\sin
^{2}\theta ^{*}d\varphi ^{*2}  \nonumber
\end{eqnarray}
or

\QTP{Body Math}
\begin{eqnarray}
ds^{2} &=&dt^{*2}-dr^{*2}-2a\sin ^{2}\theta ^{*}dr^{*}d\varphi ^{*}
\label{flat2} \\
&&-(r^{*2}+a^{2}\cos ^{2}\theta ^{*})d\theta ^{*2}-(a^{2}+r^{*2})\sin
^{2}\theta ^{*}d\varphi ^{*2}.  \nonumber
\end{eqnarray}

We will be interested in this Kerr-form of the flat metric and try to see
how the mysterious transformation, Eq.(\ref{complex-transformation1}), that
led to this form of the metric arises in a natural manner from the complex
translations, Eq.(\ref{CompTrans}), in Minkowski space, described earlier in
the text. The question we will answer is; how does the expression, (\ref
{complex-transformation1}),

\begin{equation}
r_{c}=r^{*}+ia\cos \theta ^{*}  \label{*}
\end{equation}
arise naturally and how is it used in a standard way to obtain Eq.(\ref
{flat2})?

[The origin of the second of the complex transformations, Eq.(\ref
{complex-transformation2}), assuming the first, presents no problem since $%
t^{*}=t,$ implies that, from $u_{c}=t-r$ and $u^{*}=t^{*}-r^{*}$ that 
\begin{equation}
u_{c}=t-r=t^{*}-r^{*}-ia\cos \theta =u^{*}-ia\cos \theta ^{*}.]
\end{equation}

We begin with the flat metric

\begin{equation}
ds^{2}=dt^{2}-dr^{2}-r^{2}(d\theta ^{2}+\sin ^{2}\theta d\varphi ^{2})
\label{flat}
\end{equation}
and look for the ordinary real coordinate transformation to the Kerr-form,
Eq.(\ref{flat2}),

\QTP{Body Math}
\begin{eqnarray}
ds^{2} &=&dt^{*2}-dr^{*2}-2a\sin ^{2}\theta ^{*}dr^{*}d\varphi ^{*}
\label{Kerr} \\
&&-(a^{2}\cos ^{2}\theta ^{*2}+r^{*2})d\theta ^{*2}-(r^{*2}+a^{2})\sin
^{2}\theta ^{*}d\varphi ^{*2}.  \nonumber
\end{eqnarray}

\QTP{Body Math}
\qquad From the complex translation,

\QTP{Body Math}
\begin{eqnarray}
\widehat{t} &=&t,\quad \widehat{x}=x,\quad \widehat{y}=y, \\
\widehat{z} &=&z-ia
\end{eqnarray}

\QTP{Body Math}
we have

\QTP{Body Math}
\begin{eqnarray}
r_{c}^{2} &\equiv &\widehat{x}^{2}+\widehat{y}^{2}+\widehat{z}
^{2}=x^{2}+y^{2}+z^{2}-a^{2}-2iaz  \label{r'} \\
r_{c}^{2} &=&r^{2}-a^{2}-2iaz=r^{2}-a^{2}-2iar\cos \theta ,  \label{r'^2} \\
r^{2} &=&x^{2}+y^{2}+z^{2},\quad z=r\cos \theta .
\end{eqnarray}

\QTP{Body Math}
Since $r_{c}^{2}$ is complex we can write, with real ($r^{*},q),$ by
definition, 
\begin{equation}
r_{c}=r^{*}+iq,
\end{equation}
which when substituted into Eq.(\ref{r'^2}) yields

\QTP{Body Math}
\begin{equation}
r^{*2}-q^{2}+2ir^{*}q=r^{2}-a^{2}-2iar\cos \theta
\end{equation}
or 
\begin{eqnarray}
r^{2} &=&r^{*2}-q^{2}+a^{2}  \label{real} \\
\cos \theta &=&-\frac{r^{*}q}{a\sqrt{r^{*2}-q^{2}+a^{2}}},  \label{imag} \\
\quad \varphi &=&\varphi .
\end{eqnarray}

\QTP{Body Math}
This is interpreted as the coordinate transformation between the polar
coordinates $(r,\theta ,\varphi )$ and $(r^{*},q,\varphi )$ [which turns out
to be a form of spheroidal coordinates\cite{MF}]. An alternate form of
spheroidal coordinates arises when we take

\QTP{Body Math}
\begin{equation}
q=a\cos \theta ^{*}
\end{equation}
so that

\QTP{Body Math}
\begin{equation}
r_{c}=r^{*}+ia\cos \theta ^{*}
\end{equation}
and Eqs.(\ref{real}) and (\ref{imag}) become 
\begin{eqnarray}
r^{2} &=&r^{*2}+a^{2}\sin ^{2}\theta ^{*}  \label{real2} \\
\cos \theta &=&-\frac{r^{*}\cos \theta ^{*}}{\sqrt{r^{*2}+a^{2}\sin
^{2}\theta }},  \label{imag2} \\
\quad \varphi &=&\varphi
\end{eqnarray}
With this transformation the flat metric, Eq.(\ref{flat}), becomes

\QTP{Body Math}
\begin{eqnarray}
ds^{2} &=&dt^{*2}-\left( a^{2}\cos ^{2}\theta ^{*2}+r^{*2}\right)
(r^{*2}+a^{2})^{-1}dr^{*2} \\
&&-\left( a^{2}\cos ^{2}\theta ^{*2}+r^{*2}\right) d\theta
^{*2}-(r^{*2}+a^{2})\sin ^{2}\theta ^{*}d\varphi ^{2}.
\end{eqnarray}
Changing the coordinate $\varphi $ by 
\begin{eqnarray}
\varphi &=&\varphi ^{*}-\arctan \frac{r^{*}}{a}  \label{phi-phi*} \\
d\varphi &=&d\varphi ^{*}-a(r^{*2}+a^{2})^{-1}dr^{*}  \nonumber
\end{eqnarray}
leads directly to the flat ``Kerr metric'', Eq.(\ref{flat2}) ;

\QTP{Body Math}
\begin{eqnarray}
ds^{2} &=&dt^{2}-dr^{2}-2a\sin ^{2}\theta ^{*}drd\varphi \\
&&-(r^{*2}+a^{2}\cos ^{2}\theta ^{*})d\theta ^{*2}-(a^{2}+r^{*2})\sin
^{2}\theta ^{*}d\varphi ^{2}  \nonumber
\end{eqnarray}
obtained earlier by the ``trick''. Somehow the trick circumvents the hard
detailed work of finding and doing the coordinate transformations. \textit{%
It automatically does the transformations}, Eqs.(\ref{real2}),(\ref{imag2})
and(\ref{phi-phi*}) and thus seems still to be rather mysterious even though
it gives us how, $r_{c}=r^{*}+ia\cos \theta ^{*},$ originates.

\end{document}